\documentclass[letterpaper, 10 pt, conference]{ieeeconf}  

\IEEEoverridecommandlockouts                              

\usepackage{subcaption}
\usepackage{caption}
\usepackage{graphicx}
\usepackage{hyperref}
\usepackage{tabularx}
\usepackage{multirow}
\let\labelindent\relax
\usepackage{enumitem}
\usepackage{balance}
\usepackage{amsmath}
\usepackage[table,xcdraw]{xcolor}

\title{\LARGE \bf 
Building Rapport with Self-Disclosure in Human–Robot Interaction
}

\author{Haritha Anupindi$^{1}$, Chayan Sarkar$^{1}$, Tapas Nayak$^{1}$, and Marynel V\'azquez$^{2}$\\
$^{1}$TCS Research, India\\
$^{2}$Yale University, USA
}

\begin{document}

\maketitle
\thispagestyle{empty}
\pagestyle{empty}


\begin{abstract}
As social robots become increasingly present in everyday environments, establishing rapport with users is critical for enabling sustained and meaningful human–robot interaction. In human–human communication, self-disclosure is a well-established mechanism for building rapport; however, its role and impact in human–robot interaction (HRI) are not yet well understood. Although recent advances in large language models (LLMs) have improved the linguistic competence of conversational robots, such systems are typically deployed as reactive information providers, offering limited social reciprocity and relational depth. In this article, we investigate the role of robot self-disclosure in fostering rapport during face-to-face human–robot interaction. We propose a dialogue system, called RADIA that leverages retrieval-augmented generation (RAG) to enable a robot to retrieve and reference simulated autobiographical memories, allowing first-person self-disclosure to be incorporated naturally into ongoing conversations. We evaluate the proposed approach through an in-person user study, comparing interactions with a self-disclosing robot against a baseline non-self-disclosing robot. Our results show that robot self-disclosure significantly influences conversational dynamics and enhances users’ perceived rapport with the robot, suggesting that self-disclosure is a valuable social behavior for long-term and socially engaging human–robot interaction.

\end{abstract}

\section{Introduction}

Human–robot interaction extends beyond task completion to the formation of socially meaningful relationships between users and robots~\cite{sarkar2026tool}. Central to such interactions is the development of rapport, which shapes how users engage with social robots. Rapport in human–robot interaction (HRI) is commonly understood as a dynamic, reciprocal relationship characterized by emotional connection, mutual attentiveness, and trust~\cite{tickle1990nature}. In socially sensitive domains such as eldercare, education, and healthcare, rapport plays a critical role not only in functional assistance but also in shaping user comfort, trust, and long-term acceptance of robotic systems~\cite{zonca2021role}. Consequently, understanding how robots can participate in rapport-building processes is central to the design of socially interactive agents.

Verbal communication is a primary mechanism through which rapport is established in both human–human interaction and HRI~\cite{marin2021verbal}. Conversational cues such as responsiveness, empathy, and the sharing of personal information contribute to perceived warmth and social presence. Among these cues, self-disclosure, the act of sharing personal information, has been identified as a strong driver of interpersonal connection~\cite{rime2016self}. Social Penetration Theory further suggests that relationships deepen through reciprocal exchanges of increasingly personal information; when such reciprocity is absent or asymmetric, relational development can be hindered~\cite{carpenter2015social, wheeless1977measurement}.

\begin{figure}
    \centering
    \includegraphics[width=0.99\linewidth]{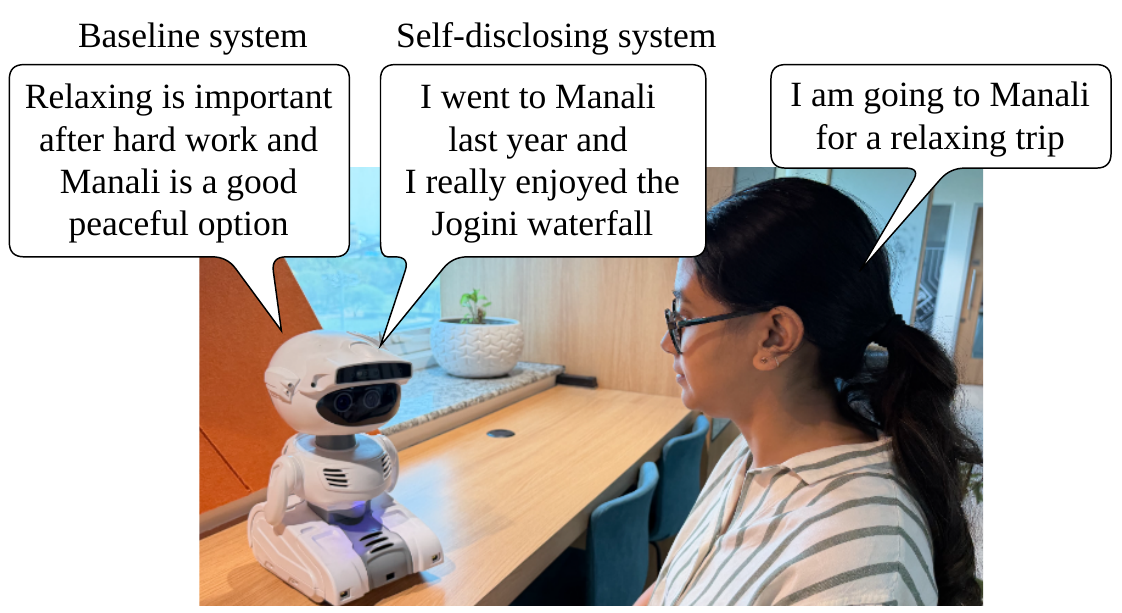}
    \caption{A user interacting with the Misty social robot with two conversation styles -- disclosure-restricted (baseline) and self-disclosing (RADIA) system.}
    \label{fig:interaction_snap}
\vspace{-2em}
\end{figure}

Despite these well-established principles, most HRI systems remain asymmetric: humans disclose personal experiences or emotions, while robots respond supportively without contributing personal information. This limitation is especially evident in contemporary LLM-based dialogue systems which, despite their linguistic fluency, are typically deployed as reactive responders rather than active social interlocutors~\cite{deng2023survey, deng2025proactive}. As a result, interaction often lacks the reciprocal structure necessary for rapport building in human conversation. While prior research has examined how conversational agents encourage or facilitate human self-disclosure~\cite{tran2025robot, zhang2025power, pujiarti2022enhancing}, relatively limited work has investigated the converse perspective: how agent-initiated self-disclosure influences conversational quality and rapport.


In this work, we address this gap by examining whether controlled, first-person self-disclosure by a conversational agent can promote reciprocal interaction and strengthen rapport in HRI. We operationalize self-disclosure as brief, first-person references to simulated autobiographical experiences~\cite{bak2014self, barak2007degree}, expressed in response to contextually relevant user utterances. To support this behavior, we introduce a rapport-aware self-disclosing agent, RADIA that leverages retrieval-augmented generation (RAG) to provide the agent access to a structured repository of autobiographical content, which can be selectively retrieved and woven into ongoing conversation as shown in Fig.~\ref{fig:interaction_snap}. Our contributions are twofold:
\begin{itemize}
    \item We propose a method for generating robot speech perceived as self-disclosure, using retrieval-augmented generation to select and contextualize personal-like content within natural conversational flow.
    \item We empirically evaluate this approach in a controlled human–robot interaction study, demonstrating that agent self-disclosure significantly influences conversational dynamics and increases perceived rapport.
\end{itemize}





\section{Related Work}
\label{sec:related}

Across prior HRI research, self-disclosure has been consistently associated with bonding, trust, and well-being; however, interactions remain largely asymmetric, with humans disclosing while agents respond supportively. The role of agent-initiated self-disclosure and its impact on conversational quality and rapport therefore remains insufficiently explored.

\subsection{Self-Disclosure in Human Communication}

Self-disclosure is a foundational mechanism in the development of interpersonal relationships in human–human interaction~\cite{collins1994self}. Social Penetration Theory (SPT)~\cite{carpenter2015social} conceptualizes relationship development as a gradual process in which conversational partners reveal increasingly personal information across dimensions of breadth and depth. A central tenet of SPT is the norm of reciprocity: personal disclosure by one individual creates a social expectation for comparable disclosure by the interlocutor, helping to maintain relational balance.

Empirical studies further underscore the importance of reciprocal self-disclosure. Sprecher \textit{et al.}~\cite{sprecher2013effects} examine disclosure roles in initial interactions and show that listening to another’s disclosure leads to greater liking and perceived closeness than disclosing alone.
Their findings highlight that relational benefits arise not merely from disclosure itself, but from reciprocal exchange. Similarly, Chen \textit{et al.}~\cite{chen2024trusting} demonstrate that immediate, turn-taking reciprocal disclosure results in higher interpersonal liking and trust than delayed or non-reciprocal patterns. This work highlights the importance of timing and symmetry in synchronous social interaction.

Taken together, these theoretical and empirical findings establish reciprocity---rather than unilateral disclosure---as a key driver of liking, trust, and relational closeness in human communication. Despite this strong foundation, the role of reciprocal self-disclosure in human–robot interaction remains largely underexplored.

\subsection{Self-Disclosure in Human–Robot Interaction}

In human–robot interaction (HRI) and conversational agent research, self-disclosure has primarily been studied as a behavior initiated by the human user rather than the agent. 


For instance, Birnbaum \textit{et al.}~\cite{birnbaum2016robots} demonstrate that a robot's supportive responses to user disclosure increases perceived intimacy and willingness to engage. Similarly, studies with virtual humans show that individuals are more willing to share sensitive information under favorable social conditions such as perceived non-judgment or rapport~\cite{stock2023tell, lucas2014s}. Together, these findings establish disclosure as a key mechanism shaping closeness and engagement in human–agent interaction.

Recent comparative work highlights limitations in reciprocal disclosure by artificial agents. While LLM-based systems produce responses high in politeness and interactional involvement, they engage in significantly less reciprocal self-disclosure than humans~\cite{hwang2025can}. Although reciprocal self-disclosure behaviors appear in constrained contexts such as small talk, agents often provide minimal first-person personal information during emotionally complex exchanges.

LLM-based conversational systems have also been explored as tools for encouraging human self-disclosure across domains including mental health and elder care. Prior work shows that conversational agents can support emotional expression and psychological well-being~\cite{li2023tell}, while agent personality design and interaction strategies can promote social engagement and disclosure among older adults~\cite{guo2025exploring, laban2022informal, noguchi2020personality}. Longitudinal evidence further indicates that disclosure toward robots increases over repeated interactions, alongside improved perceptions of social competence and well-being~\cite{laban2024building}. 
Relational agent research similarly emphasizes conversational strategies that elicit human disclosure and sustain engagement over time~\cite{bickmore2005establishing, brito2026unlocking}.

Recent work on LLM-based conversational agents explores proactivity through internal cognitive mechanisms, such as latent “inner thoughts” guiding responses. While these approaches enable autonomous behavior, they do not treat self-disclosure as a social signal for rapport building, highlighting a gap between cognitively proactive and socially reciprocal agents.~\cite{liu2025proactive}



\subsection{Rapport Measurement in HRI}

Rapport in HRI is often understood as comprising both emotional connection and interactional alignment. We adopt the Connection–Coordination Rapport (CCR) framework because it explicitly separates these two dimensions: Connection, capturing perceived warmth and relational closeness, and Coordination, capturing conversational flow and mutual responsiveness~\cite{lin2025connection, lin2026reduced}. This distinction makes CCR particularly suitable for evaluating agent self-disclosure, which is expected to influence both perceived intimacy and conversational dynamics, providing a structured and theoretically grounded measure of rapport in our experiment.

\section{Hypotheses and Experimental Design}

\begin{figure*}[h!]
    \centering
    \begin{subfigure}{0.4\textwidth}
        \includegraphics[width=\textwidth]{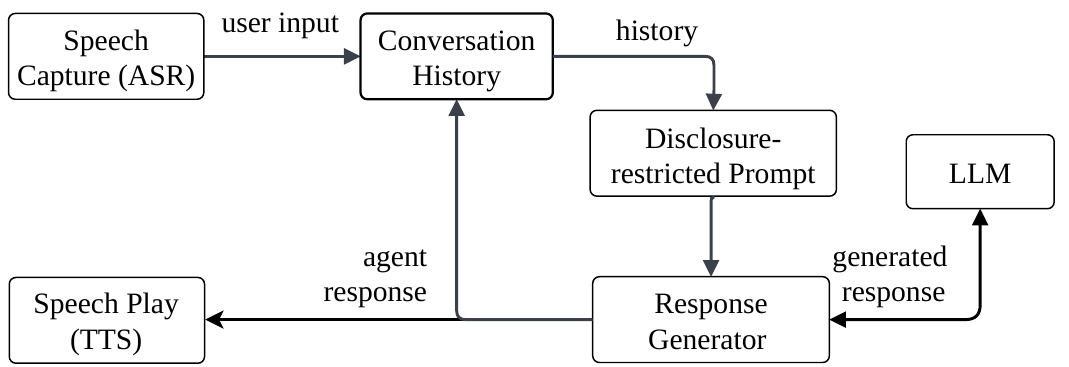}
        \caption{baseline system}
        \label{fig:sub1}
    \end{subfigure}
    \hspace{2em}
    \begin{subfigure}{0.45\textwidth}
        \includegraphics[width=\textwidth]{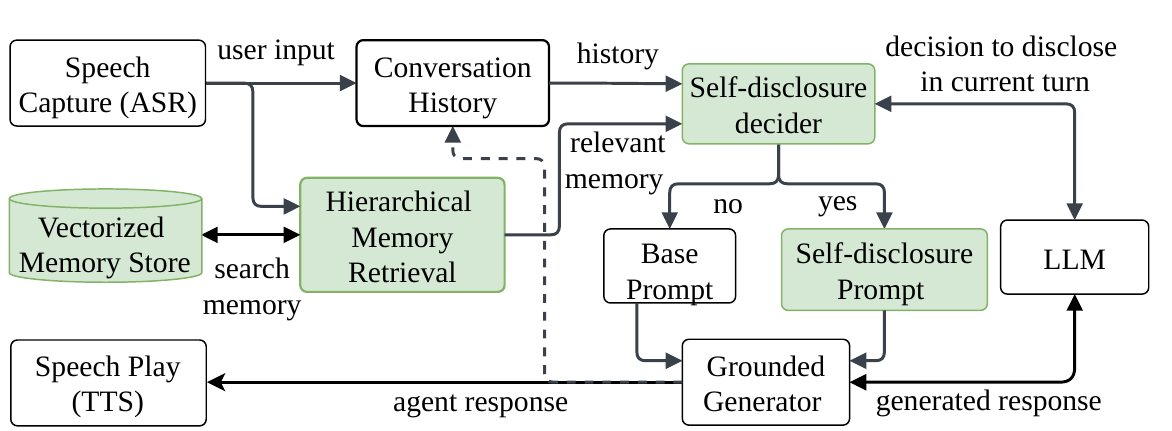}
        \caption{RADIA}
        \label{fig:sub2}
    \end{subfigure}
    \caption{System architecture of the baseline conversation system and our proposed self-disclosure enabled conversation system highlighting the additional blocks (in green).}
    \label{fig:system_design}
\vspace{-1em}
\end{figure*}






Motivated by the limited empirical understanding of how agent self-disclosure influences relational outcomes in HRI, we formulate the following hypotheses to systematically examine its effects on interaction dynamics.

\textbf{H1}: The self-disclosing agent will be scored higher in conversation quality, measured in terms of appropriateness, grammar, relevance, and content.

\textbf{H2}: The self-disclosing agent will receive higher rapport scores as measured by the CCR scale.


\textbf{H3}: Participants interacting with the self-disclosing agent will exhibit higher amounts of self-disclosure themselves.

\textbf{H4}: Participants interacting with the self-disclosing agent will demonstrate longer interaction durations.


\subsection{Independent Variables} 
\label{sec:independent_variables}

The study manipulated one primary independent variable
related to the conversational behavior of the agent.

\begin{itemize}

\item\textbf{Agent disclosure condition:}
Whether the conversational agent performed self-disclosure (RADIA) or provided neutral responses (baseline system).

\item \textbf{Interaction order (counterbalancing):}
To control for potential order effects, the sequence in which participants interacted with the two agent conditions was randomized. Participants experienced either the baseline agent followed by RADIA or the reverse order.

\end{itemize}

The study employed a within-subjects design in which each
participant interacted with both agent conditions.

    



\subsection{Dependent Variables} \label{sec:dependent_variables}

After each interaction, participants completed a short post-interaction questionnaire using 5-point Likert-scale items to assess their perceptions of the conversational agent. The questionnaire measured the following constructs:

\textbf{1. Conversation quality}: Participants evaluated the quality of the interaction across four dimensions -- appropriateness($a$), relevance($r$), grammatical correctness($g$), and content usefulness($c$), adapted from the LLM-Eval conversational evaluation framework~\cite{lin2023llm}. Scores from these 4 items were then aggregated to compute a composite conversation($CQ$) quality score as,
\begin{equation}
    CQ = \frac{1}{4}(a + r + g + c).
    \label{eq:conv_quality}
\end{equation}

\textbf{2. Rapport}: Participants evaluated rapport with the conversational agents using an eight-item scale adapted from the Reduced-Length Connection-Coordination Rapport (CCR) scale~\cite{lin2026reduced}. This instrument evaluated dimensions such as connection($r_1$), empathy($r_2$), coordination($r_3$), getting along($r_4$), respect($r_5$), smooth flow($r_6$), attentiveness($r_7$) and engagement($r_8$). Scores from these 8 items were then aggregated into a composite rapport score.
\begin{equation}
    R = \frac{1}{8} \sum_{i=1}^{8} r_i
    \label{eq:rapport_score}
\end{equation}

\textbf{3. Comparative study}: After completing both interaction conditions, participants completed an 5-item comparative questionnaire. For each item, participants indicated their preference between the two systems, or selected no noticeable difference, across aspects such as willingness to interact again, conversational balance, engagement, and overall interaction quality. These questions served as a manipulation check to determine whether participants perceived meaningful differences in conversational behavior between the systems.

\section{System Design}
\label{sec: method}

To investigate the impact of agent self-disclosure on rapport building, we designed two conversational systems: a disclosure-restricted baseline and our self-disclosing system, RADIA. Both systems share the same architecture, consisting of an automated speech recognition (ASR) model, a large language model (LLM), and a text-to-speech (TTS) model, ensuring that differences arise solely from disclosure behavior rather than model capability (Fig.~\ref{fig:system_design}). We use Whisper~\cite{radford2022robust}, gpt-4o-mini~\cite{hurst2024gpt}, and OpenVoiceTTS~\cite{qin2023openvoice} as ASR, LLM, and TTS, respectively.

\subsection{\textbf{Baseline System: Disclosure-Restricted Agent}}



The baseline agent was intentionally configured as a disclosure-restricted system to provide a controlled comparison for isolating the effects of agent self-disclosure. Because the study examined whether self-disclosure influences rapport, the baseline was constrained to avoid first-person or third-person statements that could be interpreted as disclosure. Allowing such content would blur the distinction between ordinary LLM conversational behavior and the deliberate disclosure. Thus, the baseline remained supportive, contextually appropriate, reactive, and user-centered, preserving a clear contrast with RADIA and ensuring that any observed differences could be attributed to RADIA’s controlled self-disclosure mechanism. As illustrated in Fig.~\ref{fig:sub1}, the system consists of the following components:

\subsubsection{{Speech interaction layer(ASR and TTS)}}
User speech is captured through an ASR module and transcribed into text. After response generation, a TTS module converts the agent’s reply into speech, enabling real-time spoken interaction.

\subsubsection{{Conversation history}}
Each interaction turn is appended to a continuously maintained conversation history buffer, preserving dialogue continuity and contextual grounding across turns.

\subsubsection{{Disclosure-restricted prompt}}
All responses are generated using a fixed conversational prompt that explicitly prohibits autobiographical references, experiential statements, or personal memories. The agent remains supportive and contextually responsive while maintaining a strictly non-reciprocal interaction style.

\subsubsection{{LLM response generation}}
The LLM~\cite{hurst2024gpt} generates the final response using the disclosure-restricted prompt. The response is then appended to the conversation history and converted to speech via the TTS module for delivery.

\subsection{\textbf{RADIA: \underline{R}apport-\underline{a}ware Self-\underline{Di}sclosing \underline{A}gent}}

RADIA is a memory-grounded conversational agent that enables contextually appropriate reciprocal self-disclosure. Instead of remaining purely reactive, the agent selectively introduces autobiographical information when relevant to the ongoing dialogue, supporting more socially participatory interaction. Both RADIA and the baseline system share the same speech interaction layer (ASR and TTS) and conversation history buffer to ensure comparable interaction conditions. As illustrated in ~\ref{fig:sub2}, the remaining components specific to RADIA are described below.






\subsubsection{{Vectorized memory store}}

The agent maintains a curated collection of autobiographical memories organized by semantic topics (e.g., hobbies, travel, etc). These user-constructed narratives simulate personal experiences that the agent may reference during conversation.

Memories are represented hierarchically, comprising topical categories, experience-level paragraphs, and individual disclosure sentences. To enable semantic retrieval, all memory content is encoded using the DistilRoBERTa\footnote{https://huggingface.co/sentence-transformers/all-distilroberta-v1\label{foot_note}} sentence embedding model, producing dense representations that capture semantic meaning beyond lexical similarity.

During initialization, embeddings are computed at two levels: (1) topic embeddings representing coarse semantic domains, and (2) memory embeddings generated for both paragraph-level experiences and sentence-level disclosure units, enabling fine-grained retrieval.

\subsubsection{{Hierarchical memory retrieval}}

Upon receiving a user utterance, the system performs hierarchical semantic retrieval to identify conversationally relevant memories.

\paragraph{Topic inference}

The user utterance is encoded using the DistilRoBERTa~\footref{foot_note} embedding model to maintain consistency with stored memory representations. The resulting embedding is compared with topic embeddings using cosine similarity, and the highest-scoring topic exceeding a predefined threshold is selected as the current conversational domain. This stage serves as a coarse semantic filter, restricting retrieval to a relevant region of the memory space, thereby improving computational efficiency and preventing unrelated disclosures.

\paragraph{Paragraph and sentence-level retrieval}

Retrieval then proceeds hierarchically within the selected topic. Paragraph-level experience embeddings are ranked by cosine similarity to identify the most contextually aligned autobiographical experience. Subsequently, sentences within the selected paragraph are compared against the user embedding, and the highest-ranked sentence is returned as the candidate disclosure memory. This multi-stage process ensures efficient search, topical coherence, and fine-grained contextual relevance. Importantly, retrieval only produces a candidate memory; whether disclosure occurs is determined in a later decision stage.

\subsubsection{{Self-disclosure decider}}

A key component of RADIA is a dedicated self-disclosure decider, implemented using the same LLM~\cite{hurst2024gpt} as the response generator, which determines whether a retrieved candidate memory should be revealed during the current conversational turn. This gating mechanism operationalizes principles from Social Penetration Theory~\cite{carpenter2015social}, ensuring disclosures remain gradual, contextually appropriate, and non-intrusive. The module considers the current user utterance, conversation history, and retrieved memory, producing a binary outcome:
\begin{itemize}
    \item \textbf{Yes}: the memory is forwarded to the self-disclosure prompt,
    \item \textbf{No}: response generation proceeds using the base prompt.
\end{itemize}

The resulting decision directly determines the prompt construction
strategy described in the following section.

\subsubsection{{Prompt construction and grounded generation}}

Based on the disclosure decision, the system dynamically constructs one of two generation prompts.

\paragraph{Self-disclosure prompt} The retrieved candidate memory is incorporated as the agent’s experiential context, instructing the generator to naturally integrate this memory into the response while maintaining conversational coherence and relevance.

\paragraph{Base prompt} A standard conversational prompt that guides response generation without referencing any personal or experiential content.

\subsubsection{{Grounded LLM response generation}}

The LLM~\cite{hurst2024gpt} generates the final response using the constructed prompt. The response is then appended to the conversation history and converted to speech via the TTS module for delivery.

\section{Interaction Design}

    
    
    
    \begin{figure*}[t]
\centering
\includegraphics[width=\textwidth, height=0.38\textheight]{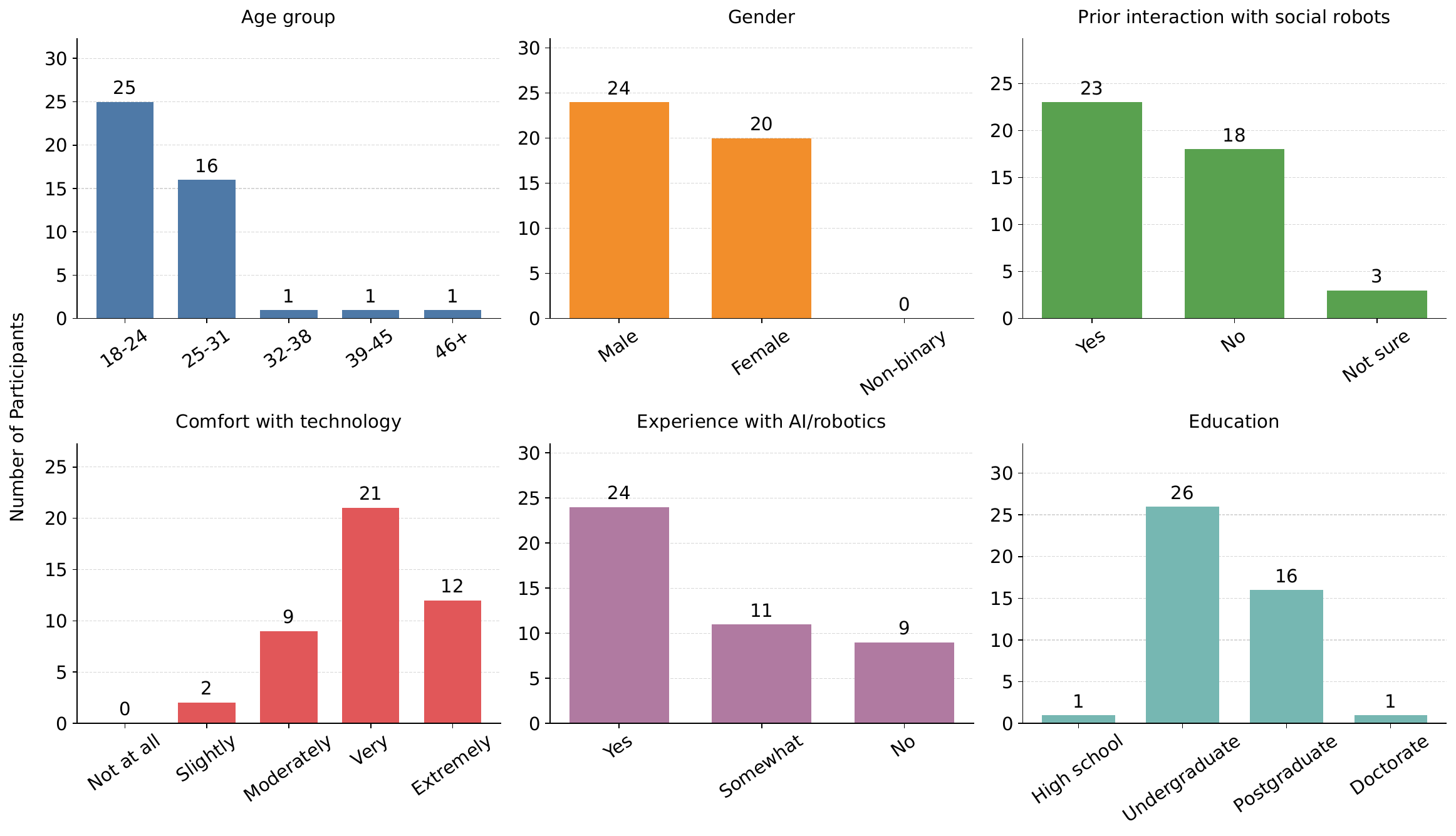}
\caption{Demographic characteristics of the study participants (N=44), showing distributions across age group, gender, prior interaction with social robots, comfort with technology, experience with AI/robotics, and educational background.}
\label{fig:demo_graph}
\end{figure*}

\subsection{Study Procedure}

We designed the experiement such that each participant engaged in two separate conversational sessions with the social robot, Misty. Each session corresponded to one experimental system, either the baseline system or RADIA. The order of system exposure was counterbalanced across participants to mitigate potential order effects.

\subsection{Interaction Guidelines}

Participants were instructed to interact freely with the robot without predefined tasks or time limits. This open-ended setup allowed conversation length to emerge naturally, mirroring real-world social exchanges where interaction duration reflects voluntary engagement.

\subsection{Post-Interaction Evaluation}

Immediately following each interaction, participants completed a post-interaction questionnaire assessing their subjective experience with the system. The questionnaire measured perceived conversation quality and rapport using Likert-scale responses, along with comparative questions designed to evaluate the study hypotheses (see Section~\ref{sec:dependent_variables} for measurement details).

\subsection{Session Duration}

The duration of the session varied between participants depending on the flow of the interaction, with all sessions completed in approximately 25 minutes.
\section{Evaluation}

\subsection{User Demographics}
A total of 44 individuals (24 male and 20 female) took part in the study. Participants were recruited through informal word-of-mouth methods, including personal networks and professional contacts. Participation was entirely voluntary, and no monetary or material compensation was provided. Demographic distribution of the participants, including age distribution, educational background, and self-reported comfort with technology, are illustrated in Fig.~\ref{fig:demo_graph}.

\subsection{Statistical Analysis Overview}

All analyses were conducted using linear mixed-effects models (LMMs) fitted with Restricted Maximum Likelihood (REML). Mixed-effects modeling was selected because each participant interacted with both systems, producing repeated measures that violate independence assumptions of traditional linear models. The models estimated fixed experimental effects while accounting for participant-level variability through random effects. Fixed effects included System (baseline, RADIA), Order (baseline-first vs. RADIA-first), and their interaction, with Participant ID modeled as a random effect.

Perceptual Likert-scale measures were analyzed using LMMs, whereas comparative categorical responses were analyzed using Pearson’s chi-square tests or Fisher’s Exact tests when expected cell counts were sparse. No significant association between response patterns and interaction order was observed ($\chi^2$ test: p = 0.12; Fisher’s Exact tests were also non-significant).

\subsection{Perceptual Measures}


\subsubsection{Statistical Reporting}
Statistical results are reported using standard metrics. The $F-statistic$ evaluates overall group differences as the ratio of explained to unexplained variance. $t-statistics$ assess pairwise mean differences by indicating how many standard deviations the difference between means is from zero. The mean ($M$) represents the average participant rating, and the standard error ($SE$) indicates the precision of that estimate. The $p-value$ reflects the probability that observed results occurred by chance, with smaller values indicating statistical significance. Practical impact is quantified using Cohen’s d (effect size), where larger values indicate stronger effects. Results are additionally reported with 95\% confidence intervals (CI), representing the range within which the true population effect is expected to lie with 95\% confidence.

\subsubsection{\textbf{Manipulation Check: Perceived Self-Disclosure and Dialogic Interaction}}

\begin{figure}
    \centering
    \includegraphics[width=0.9\linewidth]{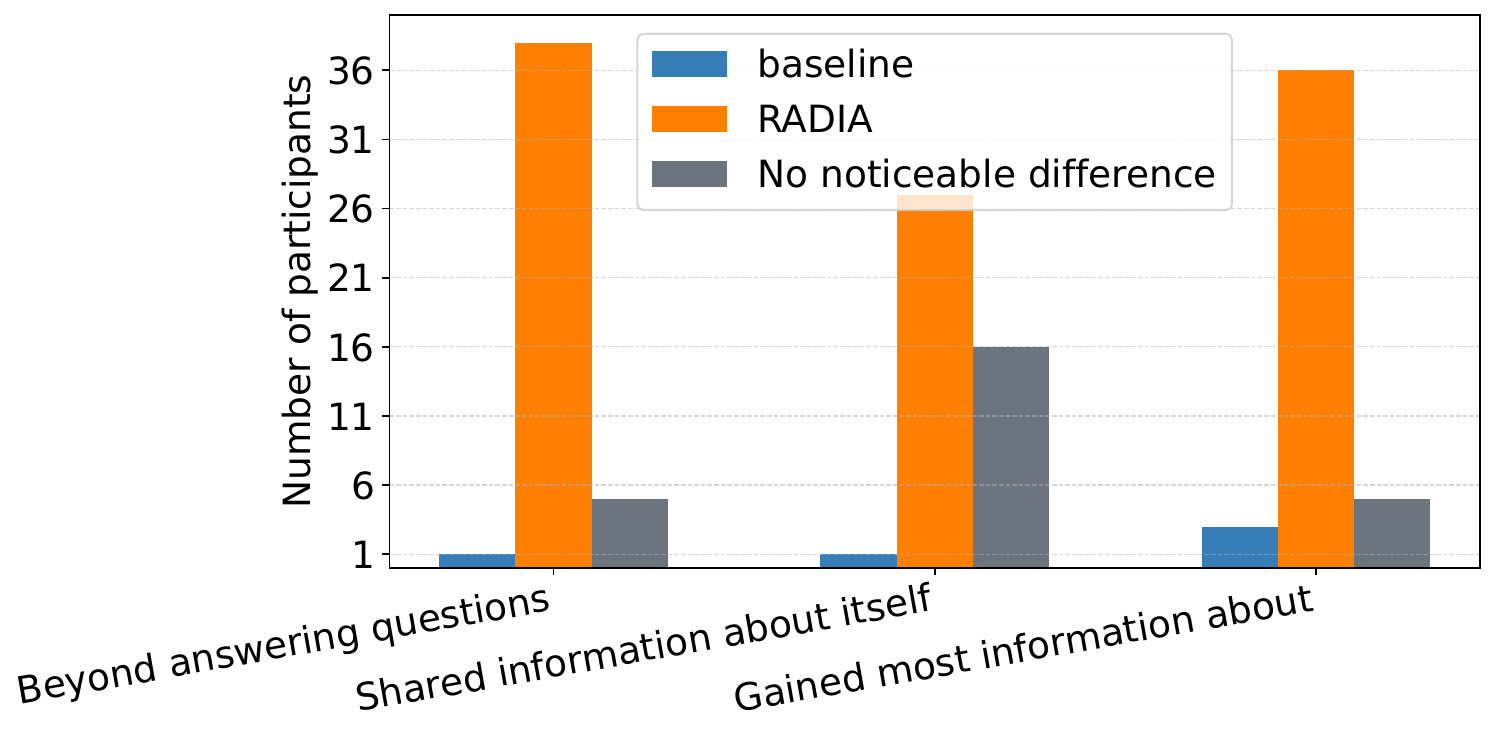}
    \caption{Validation of RADIA'S interaction characteristics.}
    \label{fig:rag_validation}
    \vspace{-1em}
\end{figure}

To verify that participants perceived behavioral differences between the systems, comparative responses assessing self-disclosure and dialogic interaction were analyzed as shown in Fig.~\ref{fig:rag_validation}.

Across all questions, participants predominantly identified RADIA as exhibiting the intended conversational behaviors. RADIA was perceived as going beyond just answering questions (86.36\%), sharing information about itself during conversation (61.36\%), and supporting reciprocal dialogue in which both sides contributed (81.82\%).

Due to sparse cell counts when responses were grouped by interaction order, Fisher’s Exact tests were conducted instead of chi-square analyses. Results indicated no significant association between response patterns and interaction order, suggesting that these perceptions were not influenced by presentation sequence.

Overall, these findings confirm that participants reliably recognized RADIA’s self-disclosing and dialogic behavior, validating the experimental manipulation prior to perceptual outcome analyses.

\subsubsection{\textbf{Conversation quality}}

\begin{figure}[t]
    \centering
    \includegraphics[width=0.99\linewidth]{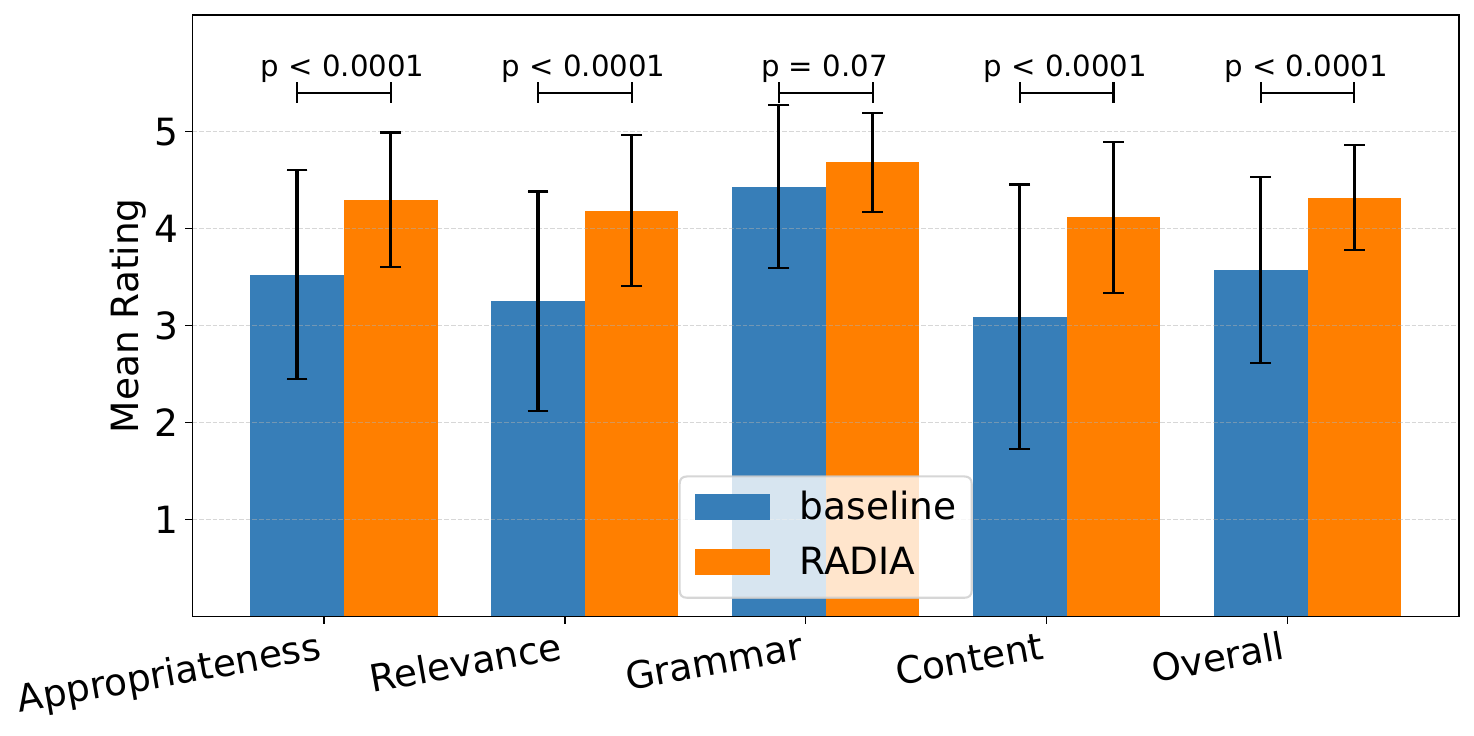}
    \caption{Conversation quality across systems (H1).}
    \label{fig:h1_visualization}
    \vspace{-1em}
\end{figure}

\begin{figure*}[h!]
    \centering
    \begin{subfigure}{0.32\textwidth}
        \includegraphics[width=\textwidth]{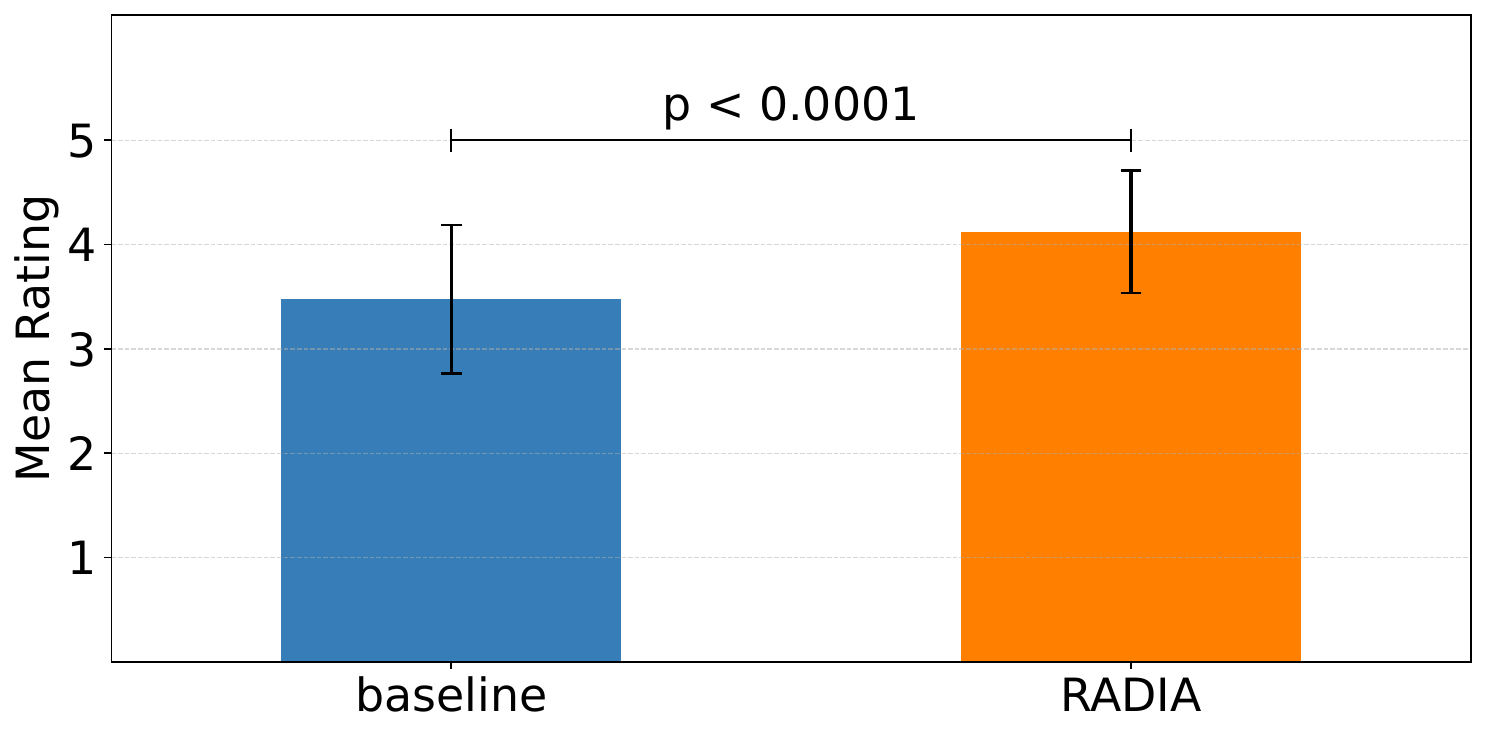}
        \caption{rapport score (H2)}
        \label{fig:h2_visualization}
    \end{subfigure}
    \begin{subfigure}{0.32\textwidth}
        \includegraphics[width=\textwidth]{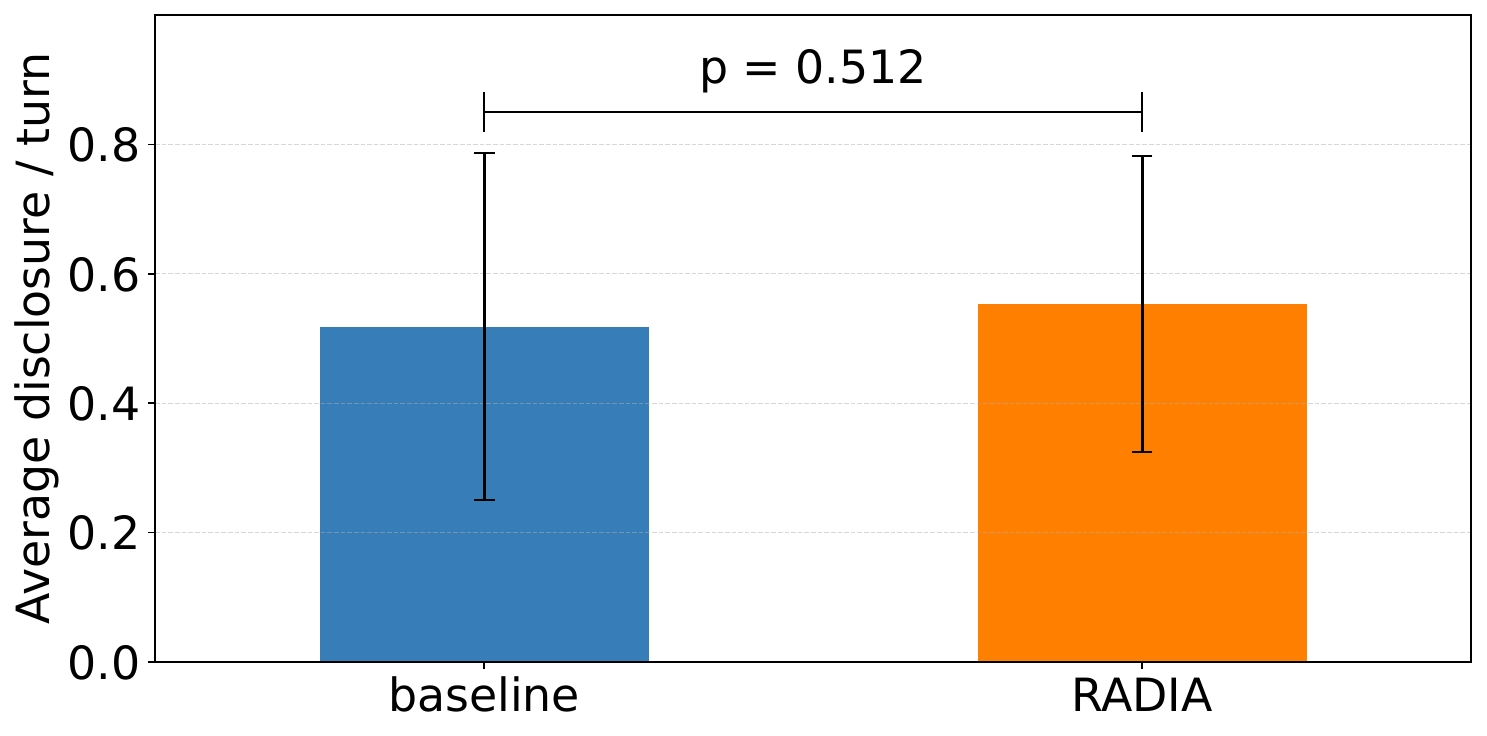}
        \caption{average disclosure (H3)}
        \label{fig:disclosure_rate}
    \end{subfigure}
    \begin{subfigure}{0.32\textwidth}
        \includegraphics[width=\textwidth]{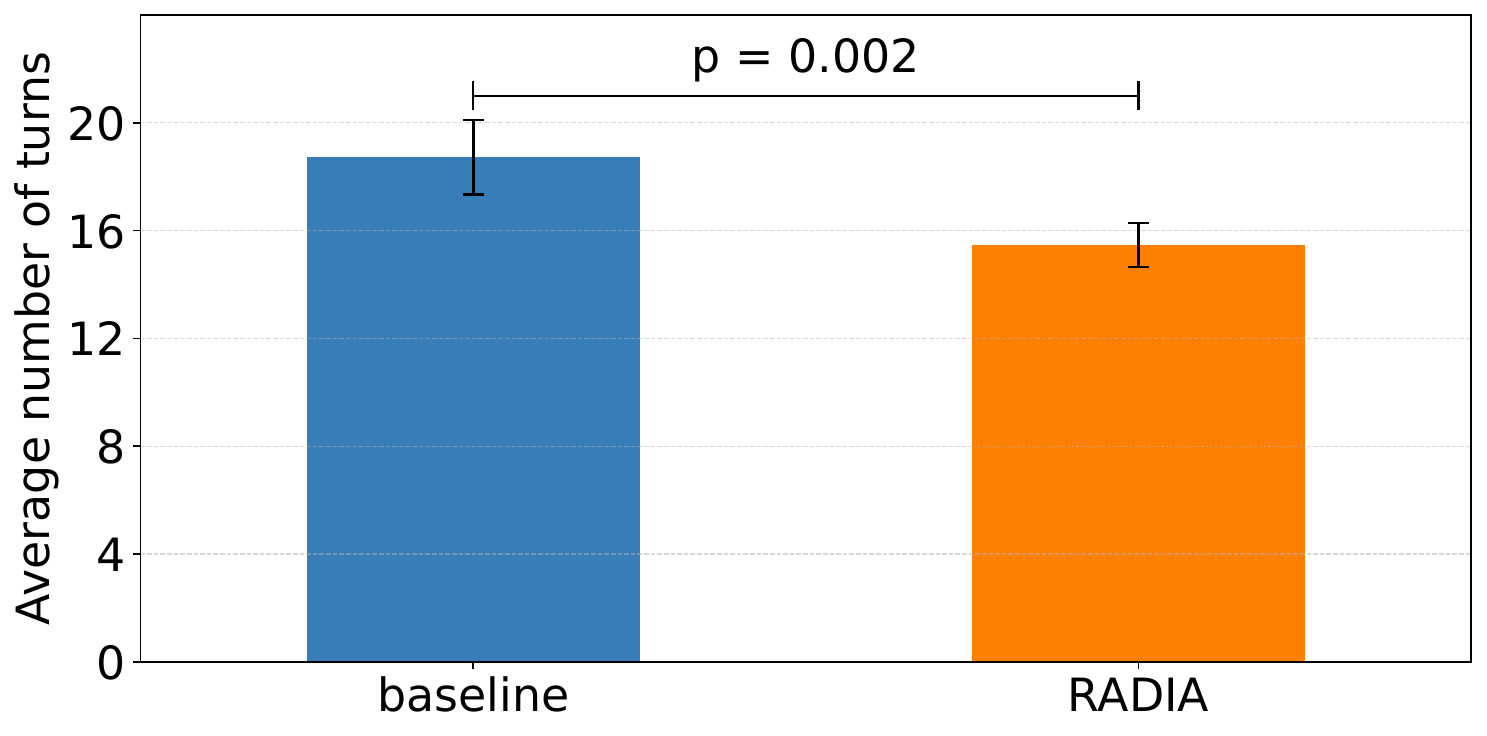}
        \caption{interaction duration (H4)}
        \label{fig:num_turns}
    \end{subfigure}
    
    \caption{Comparison of rapport score (H2), average user disclosure per turn (H3), and interaction duration in terms of user's interactions turn (H4) between baseline and our proposed system (RADIA).}
    \label{fig:hypotheses}
\vspace{-1em}
\end{figure*}

\renewcommand{\arraystretch}{1.3}

\begin{table}[t]
\resizebox{\columnwidth}{!}{%
\begin{tabular}{|l|c|c|c|c|}
\hline
\textit{\textbf{Measure}} & \textit{\textbf{t}} & \textit{\textbf{p}} & \textit{\textbf{\begin{tabular}[c]{@{}c@{}}Cohen's \\ d\end{tabular}}} & \textit{\textbf{95\% CI}} \\ \hline
\begin{tabular}[c]{@{}l@{}}Conversation \\ Quality (H1)\end{tabular} & 5.13 & \textless 0.0001 & 0.77 & \begin{tabular}[c]{@{}c@{}}{[}0.45, \\ 1.04{]}\end{tabular} \\ \hline
\begin{tabular}[c]{@{}l@{}}Rapport\\ Score (H2)\end{tabular} & 5.26 & \textless 0.0001 & 0.79 & \begin{tabular}[c]{@{}c@{}}{[}0.40, \\ 0.90{]}\end{tabular} \\ \hline
\end{tabular}%
}
\caption{Inferential statistics for conversation quality and rapport scores between System A and System B.}
\label{tab:CQR}
\end{table}

Conversation quality was evaluated using 5 point Likert-scale items as in equation~\eqref{eq:conv_quality}.




\paragraph{Item-Level Effects}
Linear mixed-model analyses revealed significant effects of system condition for three dimensions of conversation quality.

\paragraph{Appropriateness}
A significant main effect of the system was observed, $F(1,42)=25.56$, $p<0.0001$.
A post-hoc t-test indicated that RADIA ($M=4.30$, $SE=0.11$) was perceived as significantly more appropriate than the baseline system ($M=3.52$, $SE=0.16$).

\paragraph{Relevance}
A significant main effect of the system was observed, $F(1,42)=22.91$, $p<0.0001$. A post-hoc t-test indicated that RADIA ($M=4.18$, $SE=0.12$) was perceived as significantly more relevant than the baseline system ($M=3.26$, $SE=0.17$).

\paragraph{Grammatical Correctness}
The linear mixed model indicated no significant main effects, suggesting comparable grammatical performance between both systems. But, this is expected as LLMs have perfected this and also, we used the same LLM for both the system.

\begin{figure}[t]
    \centering
    \includegraphics[width=0.9\linewidth]{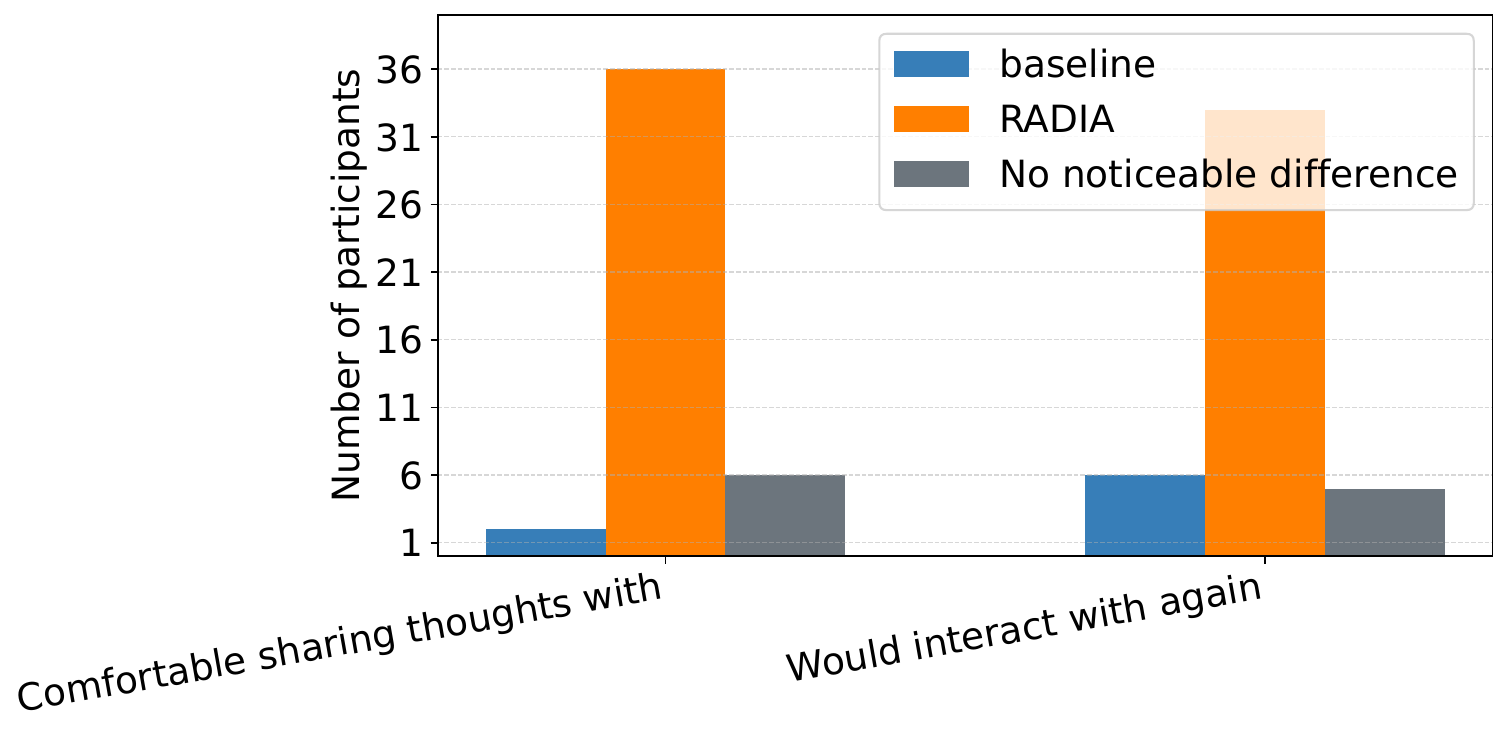}
    \caption{Overall preference across systems.}
    \label{fig:comparative_preference}
    \vspace{-2em}
\end{figure}

\paragraph{Content Usefulness}
A significant main effect of the system was observed, $F(1,42)=28.02$, $p<0.0001$. A post-hoc t-test indicated that the content of RADIA ($M=4.11$, $SE=0.12$) was perceived as significantly more useful than the baseline system ($M=3.09$, $SE=0.21$).

\paragraph{Overall Conversation Quality}
Analysis of the aggregated conversation quality score as calculated in Eq.~\eqref{eq:conv_quality} revealed a significant effect of System, $t(43)=5.13$, $p<0.0001$. As illustrated in Fig.~\ref{fig:h1_visualization}, participants rated RADIA higher overall than the baseline system. The effect size was large, with Cohen’s $d = 0.77$ and a 95\% confidence interval of $[0.45,\,1.04]$ as shown in Table~\ref{tab:CQR}, indicating a substantial improvement in perceived conversation quality for RADIA.

\paragraph{Summary (H1)}
Even though there is no significant difference between the two system based on grammatical correctness, H1 is supported based on the overall score. This demonstrates that the self-disclosing agent significantly improves perceived conversation quality, particularly through gains in appropriateness, relevance, and content usefulness. 


\subsubsection{\textbf{Rapport}}

Rapport was evaluated using 5 point Likert-scale items from the reduced-length Connection--Coordination Rapport (CCR) scale as in equation~\eqref{eq:rapport_score}.


A significant main effect of System was observed, $F(1,42)=28.36$, $p<0.0001$. A post-hoc t-test indicated that RADIA ($M=4.13$, $SE=0.09$) was rated higher on the rapport scale than the baseline system ($M=3.47$, $SE=0.11$).

\paragraph{Overall Rapport}
Analysis of the aggregated rapport score as calculated in Eq.~\eqref{eq:rapport_score} revealed a significant effect of System, $t(43)=5.26$, $p<0.0001$. As illustrated in Fig.~\ref{fig:h2_visualization}, participants rated RADIA higher overall than the baseline system. The effect size was large, with Cohen’s $d = 0.79$ and a 95\% confidence interval of $[0.40,\,0.90]$ as shown in Table~\ref{tab:CQR}, indicating a substantial improvement in perceived rapport for RADIA.

\paragraph{Summary (H2)}
These results support H2, demonstrating that RADIA significantly improves perceived rapport.

\subsubsection{\textbf{User Self-Disclosure Rate}}

User self-disclosure rate was operationalized as the average number of disclosure per turns across users. 

\paragraph{Overall Participant Disclosure Effect}
Analysis revealed no significant main effect of system on user disclosure rate ($p = 0.512$), as illustrated in Fig.~\ref{fig:disclosure_rate}. Participants exhibited comparable disclosure levels when interacting with both systems, with the baseline system showing a mean disclosure rate of $M = 0.52$, $SD = 0.27$ and RADIA showing $M = 0.55$, $SD = 0.23$. Although disclosure rate was numerically higher for RADIA, the difference was not statistically significant.

\paragraph{Summary (H3)}

Although H3 was not supported, this result suggests a boundary condition rather than a failure of RADIA. RADIA was perceived as more self-disclosing and dialogic, but this did not immediately increase users’ disclosure frequency. This may be because user disclosure is a conservative behavioral measure in short first-time robot interactions. The numerical trend was still in the expected direction, suggesting that RADIA may improve perceived rapport, while behavioral disclosure effects may require longer multi-session interactions.

\subsubsection{\textbf{Interaction Duration (Number of Turns)}}

Interaction duration was operationalized as the number of conversational turns by the participant. Please note, the agent reciprocated equal number of turns as well.

\paragraph{Overall Interaction Duration}
Analysis revealed a significant main effect of System on interaction duration ($p = 0.00204$), as illustrated in Fig.~\ref{fig:num_turns}. Participants engaged in longer conversations with the baseline system ($M = 18.73$, $SE = 1.39$) compared to RADIA ($M = 15.46$, $SE = 0.82$).

\paragraph{Summary (H4)}

Although H4 was not supported, this result may reflect interaction efficiency rather than reduced engagement. RADIA produced higher perceived conversation quality and rapport despite fewer turns, suggesting that users may have experienced the interaction as more satisfying or complete without needing to continue for longer. Number of turns is therefore a coarse proxy for engagement: longer conversations can reflect interest, but they can also reflect clarification, repetition, or lower conversational efficiency. Thus, the shorter RADIA interactions may indicate more focused and effective exchanges.

\subsubsection{\textbf{Manipulation Check: User Comfort and Interaction Preference}}

To further validate participants’ subjective preferences between systems, we analyzed comparative responses assessing comfort in sharing personal thoughts and willingness for future interaction (Fig.~\ref{fig:comparative_preference}).

For perceived comfort in sharing personal thoughts, 75\% of the participants favored RADIA. Similarly, when asked which system they would prefer to interact with again, 88.64\% participants favored RADIA.

These results indicate a clear participant preference for RADIA, suggesting that agent self-disclosure not only improved perceived interaction quality but also increased users’ comfort and willingness for continued engagement.

\section{Discussion}



This work investigated how controlled agent self-disclosure enabled through RADIA influences rapport formation in human--robot interaction. By incorporating memory-grounded first-person disclosures, RADIA allows the robot to move beyond reactive responses and participate socially in dialogue, with participants perceiving RADIA as more dialogic, socially present, and relationally engaged. These results position self-disclosure as a functional mechanism for rapport building. 

However, these perceptual gains did not translate into increased user disclosure or longer interaction duration. This perception--behavior dissociation suggests that rapport may first emerge at a cognitive or affective level before influencing observable behavior. Together, these findings highlight the need to distinguish subjective relational experience from behavioral metrics when evaluating social agents.


\textbf{Limitations \& Future Work.} The single-session design likely constrained the development of reciprocal disclosure, which often requires longitudinal interaction. Future research should conduct multi-session studies and implement adaptive disclosure strategies that adjust depth and timing based on real-time user cues. A further limitation concerns the ethics of simulated autobiographical memory, as users may interpret such disclosures as implying a lived history the robot can not literally possess. Future work should study how transparency about synthetic memories affects trust, rapport, and perceived deception in memory-grounded social robots.
\section{Conclusion}

This work introduced RADIA, a rapport-aware self-disclosing conversational agent that enables memory-grounded self-disclosure within human–robot interaction. By allowing the agent to reference experiential memories, RADIA moves beyond reactive dialogue toward socially participatory conversation. The findings demonstrate that contextually appropriate agent self-disclosure can substantially reshape users’ perceptions of conversational quality and rapport, highlighting the importance of reciprocity in human–robot interaction.

More broadly, this work positions self-disclosure as a practical design mechanism for socially intelligent agents. By grounding conversational behavior in structured memory and controlled disclosure decisions, RADIA paves a pathway toward agents that are not only informative but socially expressive partners. As conversational robots increasingly enter everyday environments, enabling systems to participate meaningfully in relational dialogue may be essential for fostering trust, acceptance, and long-term human–robot relationships.


\balance
\bibliographystyle{IEEEtran}
\bibliography{root}

\end{document}